# Bayesian Analysis of AR (1) model

Hossein Masoumi Karakani, University of Pretoria, South Africa

Janet van Niekerk, University of Pretoria, South Africa

Paul van Staden, University of Pretoria, South Africa

**Abstract:**

The first – order autoregressive process, AR (1), has been widely used and implemented in time series analysis. Different estimation methods have been employed in order to estimate the autoregressive parameter. This article focuses on subjective Bayesian estimation as opposed to objective Bayesian estimation and frequentist procedures. The truncated normal distribution is considered as a prior, to impose stationarity. The posterior distribution as well as the Bayes estimator are derived. A comparative study between the newly derived estimator and other existing estimation methods (frequentist) is employed in terms of simulation and real data. Furthermore, a posterior sensitivity analysis is performed based on four different priors; g prior, natural conjugate prior, Jeffreys' prior and truncated normal prior and the performance is compared in terms of Highest Posterior Density Region criterion.

## 1. Introduction

With the exponential growth of time-stamped data from social media, e-commerce and sensor systems, time series data that arise in many areas of scientific endeavor is of growing interest for extracting useful insights. There exist various models used for time series, however, the first-order autoregressive process, denoted as AR (1), is the most well – known and widely used model in time series analysis.

Different type of estimation methods for the AR (1) process have been developed and proposed in literature. Frequentist estimation methods, including method of moments estimation (MME), conditional least squares estimation (CLS), exact maximum likelihood estimation (MLE), and conditional maximum likelihood estimation (CMLE) are commonly used. To improve the current estimation procedures, specifically for small samples, a Bayesian method of estimation is considered for the AR (1) model. In general, reasons involving Bayesian approach in time series analysis are; firstly, this approach can successfully provide logical interpretation for statistical inferences in time series analysis, and secondly, results can always be updated based on assimilation of new information.

There is an extensive literature on autoregressive processes using Bayesian methods. Bayesian analysis of AR models began with the work of Zellner and Tiao (1964) who considered the AR (1) process. Zellner (1971), Box et al. (1976), Monahan (1984) and Marriott and Smith (1992), discuss the Bayesian approach to analyze the AR models. Lahiff (1980) developed a numerical algorithm to produce posterior and predictive analysis for AR (1) process. Diaz and Farah (1981) devoted a Bayesian technique for computing posterior analysis of AR process with an arbitrary order. Phillips (1991) discussed the implementation of different prior distributions to develop the posterior analysis of AR models with no stationarity assumption assumed. Broemeling and Cook (1993) implement a Bayesian approach to determine the posterior probability density function for the mean of a $p^{th}$ order AR model. Ghosh and Heo (2000) introduced a comparative study to some selected noninformative (objective) priors for the AR (1) model. Ibazizen and Fellag (2003), assumed a noninformative prior for the autoregressive parameter without considering the stationarity assumption for the AR (1) model. However, most literature considers a noninformative (objective) prior for the Bayesian analysis of AR (1) model without considering the stationarity assumption. See for example, DeJong and Whiteman (1991), Schotman and Van Dijk (1991), Sims and Uhlig (1991).

The intention of this paper is to impose the stationarity assumption by assuming a truncated normal distribution as an informative (subjective) prior and derive the closed form expressions for the posterior density function as well as Bayes estimator (BE).

It is worth noting that, the current article can be considered as the first pioneer to use the truncated normal distribution as subjective prior for the autoregressive parameter in AR (1) model. The article inspects the numerical effectiveness of using the aforementioned prior in finding the posterior distribution as well as BE and compare the performance of this newly derived estimator with other frequentist methods via comprehensive simulation studies and bias plot. Furthermore, the efficiency of this subjective prior (truncated normal) with three well – known priors are considered. They are g prior, asserted by Zellner (1983 and 1986), natural conjugate (NC) prior, asserted by Raiffa and Schlaifer (1961), and Jeffreys' prior, asserted by Jeffreys' (1961). In Section 2, the methodology from Bayesian perspective will be explained. Furthermore, the newly derived posterior density function as well the Bayes estimator will be discussed. Section 3 contains the comparative study in terms of simulation algorithm, comparison between newly derived BE and other well – known frequentist estimation methods in terms of different sample sizes and bias plot, posterior distribution efficiency of the truncated normal distribution with g, natural conjugate and Jeffreys' priors in terms of HDP. Finally, Section 4 conclude the implementation of newly derived results into a real dataset.

## 2. Methodology

In this section, the posterior distribution as well as the BE for the AR (1) autoregressive parameter are derived. Derivations of posterior densities depend on the likelihood function and the form of the prior distribution which are both defined hereafter.

A sequence of random variables $\{Y_t: t \geq 1\}$ defined by:

$$y_t = c + \emptyset y_{t-1} + \varepsilon_t \qquad -1 < \emptyset < 1 \qquad (1)$$

is a stationary AR (1) process (Box et al., 1970), where $y_t$ is the $t^{th}$ time series observation, $t = 1, 2, \ldots, n$, $\{\varepsilon_t\}$ is the white noise process, c is a constant term and $\emptyset$ is the autoregressive parameter.

Throughout this paper we assume that $\{\varepsilon_t\}$ is a sequence of mutually independent identically (i.i.d) distributed with a Gaussian (normal) distribution that has a mean of 0 and a common variance $\sigma_\varepsilon^2$, which will be considered a known parameter.

The conditional likelihood function of the complete sample for the AR (1) model is given by (Hamilton, 1994):

$$L(\emptyset|y_T, \ldots, y_1, \sigma_\varepsilon^2) = \prod_{t=2}^{T} \frac{1}{\sqrt{2\pi\sigma_\varepsilon^2}} \exp\left[\frac{-(y_t - \emptyset y_{t-1})^2}{2\sigma_\varepsilon^2}\right] \quad (2)$$

Since the AR (1) model is only stationary when $|\emptyset| < 1$, the truncated normal distribution is considered as a subjective prior distribution for $\emptyset$. See Johnson, Kotz and Balakrishnan (1970) for distributional details of the truncated normal distribution. The probability density function (pdf) for the prior distribution with parameters d and $\sigma_\emptyset^2$ of the autoregressive parameter, $\emptyset$, is:

$$\pi(\emptyset|\sigma_\varepsilon^2) = \frac{\frac{1}{\sqrt{2\pi\sigma_\emptyset^2}} \exp\left[\frac{-(\emptyset-d)^2}{2\sigma_\emptyset^2}\right]}{\varphi\left(\frac{b-d}{\sigma_\emptyset}\right) - \varphi\left(\frac{a-d}{\sigma_\emptyset}\right)} \quad (3)$$

where a and b are the lower and the upper truncation points and $\varphi(.)$ denotes the standard normal distribution function (CDF).

Since $-1 < \emptyset < 1$, the truncation points (a and b) can be replaced with – 1 and 1 repectively. Hence, equation (3) can be rewritten as:

$$\pi(\emptyset|\sigma_\varepsilon^2) = \frac{\frac{1}{\sqrt{2\pi\sigma_\emptyset^2}} \exp\left[\frac{-(\emptyset-d)^2}{2\sigma_\emptyset^2}\right]}{\varphi\left(\frac{1-d}{\sigma_\emptyset}\right) - \varphi\left(\frac{-1-d}{\sigma_\emptyset}\right)} \quad (4)$$

The expected value and the variance of the truncated normal distribution are (Johnson, Kotz and Balakrishnan, 1970):

$$E(\emptyset) = d + \sigma_\emptyset \frac{\emptyset(\alpha) - \emptyset(\beta)}{\varphi(\beta) - \varphi(\alpha)} \quad (5)$$

and

$$VAR(\emptyset) = \sigma_\emptyset^2 \left[1 + \frac{\alpha\emptyset(\alpha) - \beta\emptyset(\beta)}{\varphi(\beta) - \varphi(\alpha)} - \left(\frac{\emptyset(\alpha) - \emptyset(\beta)}{\varphi(\beta) - \varphi(\alpha)}\right)^2\right] \quad (6)$$

respectively, where $\alpha = \frac{-1-d}{\sigma_\emptyset}$, $\beta = \frac{1-d}{\sigma_\emptyset}$ and $\varphi(.)$ is the standard normal CDF.

**Theorem 1:**

The posterior distribution for the autoregressive parameter, $\emptyset$, of the first-order autoregressive model, AR (1), with a truncated normal prior, has a truncated normal distribution with the following pdf:

$$f(\emptyset|y_T, \ldots, y_1, \sigma_\varepsilon^2, \sigma_\emptyset^2, d) = \frac{\sqrt{f} \exp\left(-\frac{f}{2}\left(\emptyset - \frac{e}{f}\right)^2\right)}{\sqrt{2\pi}\left(\varphi\left(\sqrt{f}\left(1 - \frac{e}{f}\right)\right) - \varphi\left(\sqrt{f}\left(-1 - \frac{e}{f}\right)\right)\right)} \quad (7)$$

where $e = \frac{\sum_{t=2}^{T} y_t y_{t-1}}{\sigma_\varepsilon^2} + \frac{d}{\sigma_\emptyset^2}$ and $f = \frac{\sum_{t=2}^{T} y_{t-1}^2}{\sigma_\varepsilon^2} + \frac{1}{\sigma_\emptyset^2}$.

**Proof:**

By replacing the conditional likelihood function (2) and the prior pdf (4) into Bayes' theorem, the posterior pdf is as follows:

$$f(\emptyset|y_T, \ldots, y_1, \sigma_\varepsilon^2, \sigma_\emptyset^2, d) \propto \prod_{t=2}^{T} \frac{1}{\sqrt{2\pi\sigma_\varepsilon^2}} \exp\left[\frac{-(y_t - \emptyset y_{t-1})^2}{2\sigma_\varepsilon^2}\right] \times \frac{\frac{1}{\sqrt{2\pi\sigma_\emptyset^2}}\exp\left[\frac{-(\emptyset-d)^2}{2\sigma_\emptyset^2}\right]}{\varphi\left(\frac{1-d}{\sigma_\emptyset}\right) - \varphi\left(\frac{-1-d}{\sigma_\emptyset}\right)} \quad (8)$$

Equation (8), can be rewritten as:

$$f(\emptyset|y_T, \ldots, y_1, \sigma_\varepsilon^2, \sigma_\emptyset^2, d) \propto \left(\frac{1}{\sqrt{2\pi\sigma_\varepsilon^2}}\right)^{\frac{T-1}{2}} \exp\left(\sum_{t=2}^{T}\left[\frac{-(y_t - \emptyset y_{t-1})^2}{2\sigma_\varepsilon^2}\right]\right) \times \frac{\frac{1}{\sqrt{2\pi\sigma_\emptyset^2}}\exp\left[\frac{-(\emptyset-d)^2}{2\sigma_\emptyset^2}\right]}{\varphi\left(\frac{1-d}{\sigma_\emptyset}\right) - \varphi\left(\frac{-1-d}{\sigma_\emptyset}\right)} \quad (9)$$

**Hence:**

$$f(\emptyset|y_T, \ldots, y_1, \sigma_\varepsilon^2, \sigma_\emptyset^2, d) \propto \exp\left(-\frac{e}{2}\left(\emptyset - \frac{e}{f}\right)^2\right) \quad (10)$$

where $e = \frac{\sum_{t=2}^{T} y_t y_{t-1}}{\sigma_\varepsilon^2} + \frac{d}{\sigma_\emptyset^2}$ and $f = \frac{\sum_{t=2}^{T} y_{t-1}^2}{\sigma_\varepsilon^2} + \frac{1}{\sigma_\emptyset^2}$.

Equation (10) is identified as the kernel of a truncated normal pdf (see equation (4)) and hence the posterior pdf is:

$$f(\emptyset|y_T, \ldots, y_1, \sigma_\varepsilon^2, \sigma_\emptyset^2, d) = \frac{\sqrt{f}\exp\left(-\frac{e}{2}\left(\emptyset - \frac{e}{f}\right)^2\right)}{\sqrt{2\pi}\left(\varphi\left(\sqrt{f}\left(1 - \frac{e}{f}\right)\right) - \varphi\left(\sqrt{f}\left(-1 - \frac{e}{f}\right)\right)\right)}$$

∎

Under the squared error loss function, the Bayes estimator of the autoregressive parameter, $\emptyset$, is given by the posterior mean (5) and its accuracy can be described by the posterior variance (6).

Hence, the Bayes estimator of the autoregressive parameter, $\emptyset$, will be presented in the following theorem.

**Theorem 2:**

Under the squared error loss function, the Bayes estimator for the autoregressive parameter, $\emptyset$, of the first-order autoregressive model, AR (1), under the truncated normal distribution as a prior and the squared error loss function with reference to Theorem 1 and equation (5) is:

$$\widehat{\emptyset} = \mu_1 + \sigma_1 \frac{\emptyset\left(\frac{-1-\mu_1}{\sigma_1}\right) - \emptyset\left(\frac{1-\mu_1}{\sigma_1}\right)}{\varphi\left(\frac{1-\mu_1}{\sigma_1}\right) - \varphi\left(\frac{-1-\mu_1}{\sigma_1}\right)} \quad (12)$$

where $\mu_1 = \frac{e}{f}$, $\sigma_1 = \sqrt{f^{-1}}$, $e = \frac{\sum_{t=2}^{T} y_t y_{t-1}}{\sigma_\varepsilon^2} + \frac{d}{\sigma_\emptyset^2}$, $f = \frac{\sum_{t=2}^{T} y_{t-1}^2}{\sigma_\varepsilon^2} + \frac{1}{\sigma_\emptyset^2}$ and $\varphi(.)$ is the standard normal CDF.

### 3. Comparative Studies

This section is devoted to investigate and compare the performance of the newly derived estimator (BE) with other well – known frequentist methods of estimation. This comparison will be implemented in terms of comprehensive simulation studies and bias plot. Furthermore, the studies compare the performance of the three different priors (g, natural-conjugate and Jeffreys') with truncated normal prior using different AR (1) models and different time series lengths. The sensitivity of the posterior distribution to the change in the prior used is studied. The comparative study is implemented via Highest Posterior Density Region (HPDR). All computations are performed using MATLAB 2015a.

### 3.1 Simulation Algorithm

The current simulation study deals with data generated from the AR (1) model. For estimating the hyperparameters, training sample method (Shaarawy et al., 2010) is considered. For each time series, a training sample of size either 10 or 10% of the considered time series observations is used in this consequence. Thereafter, the estimated hyperparameters are used to conduct the simulation studies. In all simulation studies the white noise variance, $\sigma_\varepsilon^2$, is set to be 1.

The simulation algorithm for obtaining the frequentist and Bayesian estimate can be described as follows:

- Arbitrary values for the autoregressive parameter, $\emptyset$, are chosen as – 0.9, - 0.5, 0, 0.5 and 0.9;
- Time series lengths are T = 30 and T = 100;
- The white noise variance $\sigma_\varepsilon^2$, is set to be 1; and
- The hyperparameters for the truncated normal distribution are chosen by implementing training sample method.

Comparison for the sensitivity and efficiency of different priors for the autoregressive parameter, starts by generating a set of 500 normal variates $\varepsilon_t$ samples for each model of length 700. For each sample, the first 200 observations are ignored to overcome the effect of the initial values. Five different time series lengths have been chosen to study the influence of the series length on the performance of different prior distributions. These lengths are 30, 50, 100, 200 and 500. Six cases of AR (1) models are considered, for which, the values of the autoregressive parameter, $\emptyset$, were $\pm 0.2, \pm 0.5$ and $\pm 0.8$, respectively. These values were chosen inside the stationarity domain to fulfil this crucial assumption for the autoregressive parameter of the AR (1) model. The comparative study depends on some criteria (HPDR) as will be discussed in detail later.

### 3.2 Frequentist vs Bayesian estimate

In this section, the newly derived BE is illustrated and evaluated against the MME, CLS, MLE and CMLE. The evaluation and the comparison will be performed in terms of numerical estimation as well as bias plot provided.

### 3.2.1 Numerical estimation

The frequentist and Bayes estimates for the autoregressive parameter, $\emptyset$, are shown in Tables 1 and 2 based on different time series lengths (T):

| $\emptyset$ | MME | CLS | MLE | CMLE | BE |
|---|---|---|---|---|---|
| $-0.9$ | $-0.9168$ | $-0.9436$ | $-0.9278$ | $-0.9437$ | $-0.9196$ |
| $-0.5$ | $-0.5109$ | $-0.6098$ | $-0.5975$ | $-0.6106$ | $-0.5102$ |
| $0$ | $0.0190$ | $0.0191$ | $0.0186$ | $0.0191$ | $0.0199$ |
| $0.5$ | $0.5229$ | $0.5279$ | $0.5123$ | $0.5267$ | $0.5042$ |
| $0.9$ | $0.8851$ | $0.9120$ | $0.8720$ | $0.8862$ | $0.9020$ |

**Table 1: Estimates based on different values for ∅, with time series lengths, T = 30, white noise variance, $\sigma_\varepsilon^2 = 1$.**

| ∅ | MME | CLS | MLE | CMLE | BE |
|---|---|---|---|---|---|
| $-0.9$ | $-0.8978$ | $-0.8989$ | $-0.8981$ | $-0.8989$ | $-0.8992$ |
| $-0.5$ | $-0.5001$ | $-0.5001$ | $-0.4997$ | $-0.5001$ | $-0.5007$ |
| $0$ | $-0.0074$ | $-0.0074$ | $-0.0074$ | $0.0074$ | $-0.0074$ |
| $0.5$ | $0.5064$ | $0.5068$ | $0.5064$ | $0.5068$ | $0.5076$ |
| $0.9$ | $0.8925$ | $0.8933$ | $0.8927$ | $0.8925$ | $0.8953$ |

**Table 2: Estimates based on different values for ∅, with time series lengths, T = 100, white noise variance, $\sigma_\varepsilon^2 = 1$.**

In terms of bias, according to Tables 1 and 2, the BE performs the best for small samples for most of the cases and is comparative for large samples.

### 3.2.2 Bias Plot

For further investigation, the simulation study is repeated 10 times. For each estimate in each replication the absolute bias is calculated and displayed graphically in Figure 1 for ∅ = 0.5.

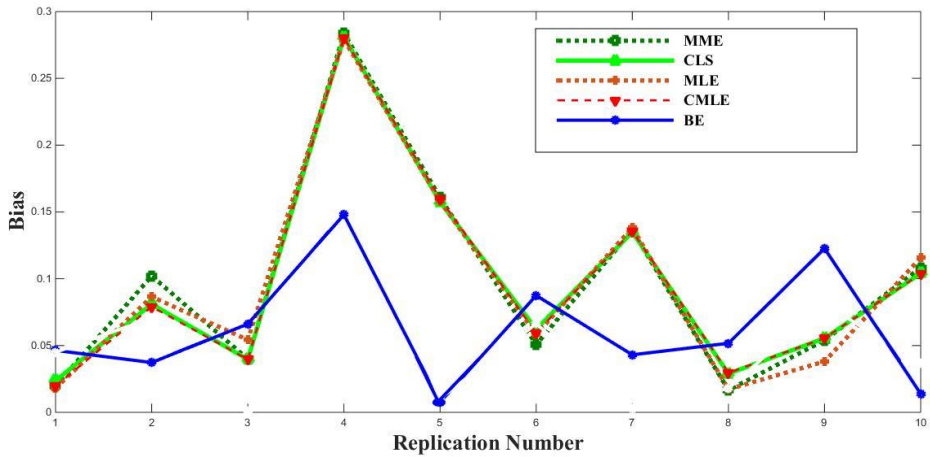

**Figure 1: Bias plot for the autoregressive parameter, ∅, estimates.**

From Figure 1, one can make a comparison between estimates based on the values of the bias. The BE (blue line) is competitive compared to other frequentist estimates. Frequentist estimates (MME, CLS, MLE and CMLE) bias levels are approximately the same.

### 3.3 Posterior Distribution Sensitivity

Various frequentist criteria are helpful to compare among prior distributions. The basic idea is to use the prior distribution to generate a posterior distribution, and investigate the frequentist properties of such resulted distribution (Yang, 1994).

An interesting tool will be used to determine the reasonable prior distribution. It is a percentage measure for the number of samples that satisfy some condition. The criterion can be explained as follows:

95% Highest Posterior Density Region (HPDR) that is defined as the region under the posterior density over the interval centered at the posterior mean with probability 95%. For each simulation, $n^*$ is defined to be the number of samples where the 95% HPDR contains the true value of the autoregressive parameter. Then, the percentage P is evaluated such that:

$$P = \frac{n^*}{500} \times 100 \qquad (13)$$

The performance of a prior is evaluated according to the value of P. In other words, for a given prior, the greater percentage indicates a higher performance of the prior to guide to a posterior that presents powerfully the autoregressive parameter.

For the posterior distribution sensitivity of AR (1) models, the stationarity assumption for the autoregressive parameter will be considered. The posterior outputs of all the proposed six AR (1) models will be studied using truncated normal (TN) prior, g prior, natural conjugate (NC) prior and Jeffreys' prior. The algorithm of the comparative analysis was implemented according to the following outlines. For each of the 500 samples; the first 30 observations used to evaluate the posterior distribution of the autoregressive parameter via four candidate priors. The posterior mean and the posterior variance of the autoregressive parameter were computed given each prior. Tracing the HPDR method, an interval centered at the posterior mean with probability 0.95 is evaluated. For each model, the percentage of samples for which the actual autoregressive parameter exists within the indicated interval was computed. Similarly, the process is repeated for the first 100, 200 and 500 observations.

The results for each of the six models are summarized throughout six tables. These tables represent the percentage (P) defined by (13) for each $n^*$. The first column represents the time series length, while each column matches the used prior distribution. The values in the cells of the table denote the percentages P.

| n | Jeff. Prior | g Prior | NC Prior | TN Prior |
|---|---|---|---|---|
| 30 | 93.4 | 94.4 | 95 | 97.2 |
| 50 | 94.5 | 94.5 | 95.2 | 96.4 |
| 100 | 94 | 94.2 | 94.5 | 94.6 |
| 200 | 95.6 | 95.4 | 95.4 | 95.8 |
| 500 | 96 | 94.8 | 94.6 | 94.4 |

**Table 3: Highest Posterior Density Region values based on different priors when the autoregressive parameter is – 0.2.**

| n | Jeff. Prior | g Prior | NC Prior | TN Prior |
|---|---|---|---|---|
| 30 | 94.4 | 94 | 95.6 | 97.6 |
| 50 | 95 | 94.6 | 95.2 | 97.2 |
| 100 | 94 | 94.6 | 94.8 | 96.2 |
| 200 | 96 | 95.8 | 95.8 | 95.8 |
| 500 | 95.4 | 95.4 | 95.2 | 95.6 |

**Table 4: Highest Posterior Density Region values based on different priors when the autoregressive parameter is 0.2.**

| n | Jeff. Prior | g Prior | NC Prior | TN Prior |
|---|---|---|---|---|
| 30 | 95.4 | 94.4 | 96 | 97.4 |
| 50 | 96.4 | 96.2 | 96.6 | 97 |
| 100 | 95.4 | 94.2 | 94.4 | 95.8 |
| 200 | 95.8 | 96.8 | 96.4 | 96.4 |
| 500 | 96 | 95.4 | 95.2 | 95 |

**Table 5: Highest Posterior Density Region values based on different priors when the autoregressive parameter is – 0.5.**

| n | Jeff. Prior | g Prior | NC Prior | TN Prior |
|---|---|---|---|---|
| 30 | 94.4 | 94.8 | 95.8 | 97.6 |
| 50 | 94.6 | 94.4 | 95.2 | 96.4 |
| 100 | 94.2 | 94.6 | 94.6 | 94.4 |
| 200 | 95.4 | 96.2 | 96 | 96 |
| 500 | 93.8 | 94.6 | 94 | 94.2 |

**Table 6: Highest Posterior Density Region values based on different priors when the autoregressive parameter is 0.5.**

| n | Jeff. Prior | g Prior | NC Prior | TN Prior |
|---|---|---|---|---|
| 30 | 95.4 | 95.6 | 97.6 | 98.4 |
| 50 | 94.4 | 95.8 | 96.4 | 97.2 |
| 100 | 94.2 | 94.4 | 94.2 | 94.2 |
| 200 | 95.2 | 96.2 | 95.6 | 95.4 |
| 500 | 96.4 | 98.2 | 96.6 | 96.4 |

**Table 7: Highest Posterior Density Region values based on different priors when the autoregressive parameter is – 0.8.**

| n | Jeff. Prior | g Prior | NC Prior | TN Prior |
|---|---|---|---|---|
| 30 | 95.4 | 94.6 | 98.2 | 98.8 |
| 50 | 94 | 93.8 | 95.4 | 96.6 |
| 100 | 94 | 95.4 | 95 | 95.2 |
| 200 | 92.8 | 94.4 | 93.6 | 94.6 |
| 500 | 93.6 | 95.2 | 95.4 | 95 |

**Table 8: Highest Posterior Density Region values based on different priors when the autoregressive parameter is 0.8.**

Regarding the above tables, we achieve the following conclusion:

1. All priors lead to consistent posterior, in the sense that the HPDR includes the autoregressive parameter value in more than 90% of the cases at all time series lengths. However, for the small sample sizes (30 and 50) the subjective prior distribution (TN) considered in this article provides better coverage of the autoregressive parameter comparing to other 3 priors and supports the findings. Since, TN assumes the stationarity assumptions for the autoregressive parameter, therefore it is of extreme importance to consider this assumption in the prior distribution and Bayesian analysis of AR (1) model.

2. For the small sample sizes (30 and 50) truncated normal prior is highly better than other priors. In terms of moderate to large sample sizes (100, 200 and 300), truncated normal prior is competitive with other priors and all of the priors are highly better than Jeffreys' prior which appears to be less consistent at all time series lengths.
3. The goodness of each prior is insensitive to the increase of the time series length.

Hence, the above results support the use of truncated normal prior, since it assumes the stationarity assumption of the AR (1) model and it avoids the problem of estimating the hyperparameters as well. Furthermore, the truncated normal prior appears to be a good choice for small time series length (30 and 50). However, for longer time series lengths, truncated normal prior is competitive with other priors.

## 4. Blowfly data

To illustrate the achieve results of the simulation study in Section 3, the blowfly dataset from (Wei, 1990) is considered. It consists of the number of adults blowflies with balanced sex ratios kept inside a cage and given a fixed amount of food daily. The sample size is 82. A graphical representation using SAS is enclosed to describe this dataset through a time plot. Phillips – Perron test is conducted to test for stationarity assumption. Moreover, the estimation of the autoregressive parameter will be considered in terms of Bayesian and frequentist estimation methods. Posterior sensitivity is also implemented by using different priors and calculate the 95% HPDR.

### 4.1 Time series analysis

Time series analysis consists of different steps and all of these steps need to be considered in order to find a reliable time series model. In Table 9, step by step approach on time series analysis is shown and will be considered thereafter:

| **Steps** | **Explanation** |
|---|---|
| **1** | **Time series visualization** |
| **2** | **Testing for stationarity** |
| **3** | **Model identification** |
| **4** | **Estimation** |
| **5** | **Model diagnostics and residual analysis** |

**Table 9: Time series analysis steps.**

Note that all of the steps and calculations is performed using SAS 9.4.

The graphical representation of blowfly dataset which is known as time series plot, is illustrated in the following figure:

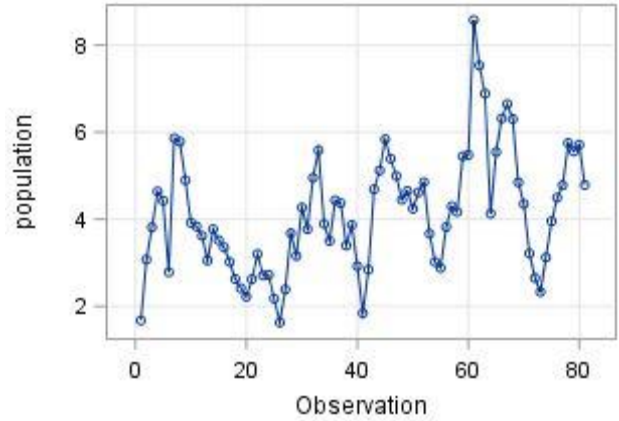

Figure 2: Time series analysis steps.

From Figure 2, the blowfly data has no cycle, trend or seasonal pattern and only irregular variation can be identified from the time series plot. Therefore, one can make a conclusion that this time series is stationary. However, more analysis is required to test for stationarity of this dataset.

The Phillips - Perron test (Phillips and Perron, 1988) is utilized to check for the stationarity assumption. The results for this test are shown in the following table:

| Lags | Rho | Pr < Rho | Tau | Pr < Tau |
|---|---|---|---|---|
| 0 | $-21.4770$ | $0.0052$ | $-3.64$ | $0.0068$ |
| 1 | $-23.1737$ | $0.0032$ | $-3.75$ | $0.0049$ |
| 2 | $-24.1534$ | $0.0024$ | $-3.81$ | $0.0041$ |
| 3 | $-23.6676$ | $0.0028$ | $-3.78$ | $0.0045$ |

**Table 10: Phillips – Perron Unit Root Tests.**

The null hypothesis of the unit root test is rejected, since all of the p – values for different lags are less than 0.01. Therefore, the null hypothesis can be rejected at 1% level of significance. As a conclusion it can be concluded that the blowfly data is stationary.

The next step is to identify the model as well as the order of the model. The minimum information criterion (MINIC) method (Hannan and Rissanen, 1982) results are shown in Table 11.

| Lags | MA 0 | MA 1 | MA 2 |
|---|---|---|---|
| AR 0 | $14.3302$ | $13.9195$ | $13.6182$ |
| AR 1 | $13.3983$ | $13.4216$ | $13.4496$ |
| AR 2 | $13.4197$ | $13.4717$ | $13.5015$ |

**Table 11: Minimum information Criterion (MINIC) results.**

From Table 11, since the minimum value is 13.3983, the chosen model is AR (1).

The next step is to estimate the autoregressive parameter for the blowfly data. The values for the hyperparameters are chosen as 0.75 and 1. Furthermore, the value for the common variance, $\sigma_\varepsilon^2$, is chosen as 1. The estimated values for the autoregressive parameter, $\emptyset$ based on the frequentist and Bayesian estimation methods are given in Table 12.

| Parameter | MME | CLS | MLE | CMLE | BE |
|---|---|---|---|---|---|
| $\emptyset$ | $0.7348$ | $0.7570$ | $0.7564$ | $0.7348$ | $0.7560$ |

**Table 12: Blowfly data estimates for ∅.**

The final step for time series analysis is to determine whether the fitted model is an adequate representation of the process that generated the observed time series. Residual analysis is useful in order to identify the validity of a model.

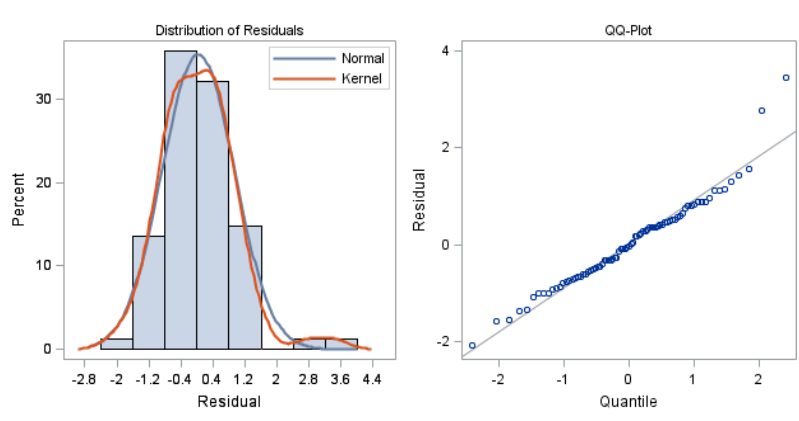

**Figure 3: Residual Normality Diagnostics for blowfly data.**

The histogram for the residuals in Figure 3 is approximately symmetric around zero. The normal density curve (kernel) added to the histogram suggest that the residuals may be normal. This is verified with the quantile – quantile (Q – Q) plot on the right hand side of Figure 3.

Various tests for normality are given in Table 13. P – values for all of these test is greater than 0.05. Therefore, the null hypothesis of normality cannot be rejected at a 5% level of significance. Hence, the residuals are normally distributed.

| Test | Statistic | P – value |
|---|---|---|
| **Kolmogorov – Smirnov** | 0.0610 | > 0.150 |
| **Cramer – von Mises** | 0.0676 | > 0.250 |
| **Anderson – Darling** | 0.5814 | 0.131 |

**Table 13: Goodness of fit Tests for Normal Distribution for blowfly data.**

## 4.2 Posterior analysis

The next step after considering time series analysis, is to demonstrate the posterior analysis that was accomplished in the previous section and compared over the four candidate priors; g prior, NC prior, Jeffreys' prior and truncated normal prior. For the blowfly dataset, the posterior mean and the 95% HPDR centered at the posterior mean are evaluated with respect to the four proposed prior distributions. The results of posterior analysis are summarized through the following table (Table 14):

| Prior | Posterior Mean | 95% HPDR |
|---|---|---|
| Jeffreys' prior | 0.7490 | [0.7350, 0.7850] |
| g prior | 0.7510 | [0.7430, 0.7930] |
| NC prior | 0.7540 | [0.7490, 0.7890] |
| TN prior | 0.7560 | [0.7490, 0.7590] |

**Table 14: Posterior Mean of ∅ and the 95% HPDR centered at the posterior mean by Prior Distribution for the blowfly data (n = 82).**

Examining the above results shows similar conclusions for the posterior analysis. The performance of the four priors based on the posterior mean values are almost the same. The unique difference is shown through the 95% HPDR that supposed to give a probability 0.95 with shortest interval. Therefore, the length of the computed interval is taken as a powerful tool to compare the performance of the priors.

Table 14 shows that the TN prior gave the shortest interval with length 0.02 comparing to (Broemeling & Cook, 1993) other priors.

## 5. Conclusion

In this paper, a subjective Bayesian analysis of the AR (1) model was done. Furthermore, the posterior sensitivity analysis was performed based on four different priors; Jeffreys' prior, g prior, natural conjugate prior and truncated normal prior. The main contributions can be summarized as follows:

- Closed form expressions for the posterior pdf and Bayes estimator (BE) are derived under the truncated normal prior.
- The newly derived BE was compared to the well – known frequentist estimation methods. It showed to be a competitive choice for all sample sizes and superior for small samples.
- In terms of posterior sensitivity based on four different priors, truncated normal distribution that was considered in this paper to assume the stationarity assumption, outperforms other priors in terms of Highest Posterior Density Region (HPDR) criterion.
- The study considered real time series example (blowfly data) to illustrate the time series analysis steps and the process of prior selection in the posterior analysis in real life. The blowfly dataset follows the AR (1) process. Posterior analysis of the blowfly data showed similar results comparing to the simulation study.